\documentclass[twocolumn,prl,aps]{revtex4}
\usepackage{amssymb}
\usepackage{epsfig,amsmath}
\usepackage{graphicx}
\usepackage{dcolumn}
\usepackage{bm}

\newcommand{\beq}{\begin{equation}}
\newcommand{\eeq}{\end{equation}}
\newcommand{\beqa}{\begin{eqnarray}}
\newcommand{\eeqa}{\end{eqnarray}} \newcommand{\lam}{\lambda}

\newcommand{\ra}{\rangle}

\def\ajp#1{{ Am.\ J.\ Phys.} {\bf #1}}

\def\jmo#1{{ J.\ Mod.\ Opt.} {\bf#1}}
\def\jpb#1{{ J.\ Phys.\ B} {\bf#1}}

\def\nat#1{{ Nature} {\bf#1}}

\def\njp#1{{ New\ J.\ Phys.} {\bf#1}}

\def\pra#1{{ Phys.\ Rev. A\/} {\bf#1}}

\def\prl#1{{ Phys.\ Rev.\ Lett.} {\bf#1}}

\def\sci#1{{ Science} {\bf#1}}
\def\rmp#1{{ Rev. \ Mod. \ Phys.} {\bf#1}}

\begin{document}

\title{Quantifying Quantum Resource Sharing}
\author{Xiao-Feng Qian$^{1,2}$}
\author{Miguel A. Alonso$^{1,3}$}
\author{J.H. Eberly$^{1,2,3}$}
\affiliation{$^{1}$Center for Coherence and Quantum Optics, University
of Rochester, Rochester, New York 14627, USA\\
$^{2}$Department of Physics \& Astronomy, University
of Rochester, Rochester, New York 14627, USA\\
$^{3}$The Institute of Optics, University of Rochester,
Rochester, NY 14627, USA}

\date{\today }


\begin{abstract}
Entanglement is a key resource of quantum science for tasks that
require it to be shared among participants. Within atomic, condensed
matter and photonic many-body systems the distribution and sharing of
entanglement is of particular importance for information processing
by progressively larger and larger quantum networks. Here we report a
singly-bipartitioned qubit entanglement inequality that applies to any
N-party qubit pure state and is completely tight. It provides the
first prescription for a direct calculation of the amount of
entanglement sharing that is possible among N qubit parties. A
geometric representation of the measure is easily visualized via
polytopes within entanglement hypercubes.
\end{abstract}

\maketitle



\noindent{\bf Introduction:} Resource sharing is a concern everywhere
in science and technology, as well as in everyday life. It always
becomes harder to allocate resources if there are more recipients,
and is still harder if there are complicated restrictions on the
allocations. These obvious considerations can become acute when the
resource has specific or even uniquely valuable qualities. A prime
example of this is many-body quantum entanglement
\cite{Horodecki-etal-09}, a property of quantum systems that is not
only desirable but necessary in order to capture quantum advantages
in applications such as randomness generation \cite{Pironio-etal-10},
cryptography \cite{Gisin-etal-02}, computing \cite{Nielsen-Chuang-00}
and network formation \cite{Kimble-08}. Despite the long-recognized
value of entanglement, it has remained unknown how to determine
either the kind or amount of qubit entanglement present in an
arbitrary many-body solid state, atomic or photonic system
\cite{Vidal-00}. More particularly, the ways that quantum
restrictions affect sharing between or among units in the system have
remained mysterious. A key obstacle is the inability to recognize,
much less catalog, all restrictions. These are open issues affecting,
for example, multi-electron atomic ionization \cite{Becker-etal-12},
multilevel coding for quantum key distribution \cite{Multi-QKD} and
multiparty teleportation \cite{MultiParty-tele, MultiEnt-tele}.

An early advance more than a decade ago used concurrence
\cite{Wootters-98} as the entanglement measure to identify the
concept of quantum monogamy (see Coffman, et al.
\cite{Coffman-etal-00}). As applied to three qubits it demonstrated
sharing or additivity by proving that the squared concurrence of
qubit $A$, with qubits $B$ and $C$ being considered as a unit, must
be greater than or equal to the sum of squared concurrences of $A$
with $B$, and $A$ with $C$, when $B$ and $C$ are considered
separately. The positive difference is known to serve as a three-body
entanglement monotone \cite{DVC-00}. The monogamy proof is symmetric
and generic, applying to any three-qubit pure state, and each qubit
is allowed to arrange its two states arbitrarily. A series of
extensions \cite{Osborne-Verstraete-06, Lohmayer-etal-06, Ou-Fan-07,
Hiroshima-etal-07, Eltschka-etal-09, Giorgi-11,Streltsov-etal-12, Bai-etal-14, Regula-etal-14} have
shown that the monogamy inequality continues to hold for $N$-qubit
states. So far, in no case has sharing been quantified.

We have been exploring \cite{Qian-Eberly-10, Qian-etal-14} influences
on quantum interactions arising from poorly known or hidden
background parties, i.e., many-body quantum systems that have
unspecified entanglements, both among themselves and with a
designated qubit of interest. We here report one of the consequences,
the discovery of a generic and completely tight inequality among one-party marginal qubit
entanglements, a symmetric linear relation restricting the
entanglements of each qubit with all of the others. It is
an inequality that applies to every qubit of any particle type in an arbitrary
many-body pure-state qubit system. As a consequence, we can report a
quantitative prescription for the amount of sharing that is possible
of such entanglements. A laboratory examination of entanglement
dynamics reported earlier \cite{Brazil} and based on an incomplete
version \cite{Qian-Eberly-10} of the new entanglement inequality,
suggests that it will be experimentally accessible. \\


\noindent{\bf {\it N}-Party Entanglement Sharing}:\ An arbitrarily
entangled $N$-qubit pure state is written
\begin{equation}
|\Psi _{1,2,...,N}\rangle
=\sum_{s_{1},...,s_{N} = 0,1} c_{s_{1},...,s_{N}}|s_{1}\rangle
...|s_{N}\rangle ,  \label{arbitrary pure state}
\end{equation}%
where $c_{s_{1},...,s_{N}}$ are normalized coefficients and $s_{j}$
takes values 0 or 1 corresponding to the two states $|0\rangle $,
$|1\rangle $ of the $j$-th qubit, with $j=1,2,3,...,N$. In this
$N$-qubit system, we compute the degree of entanglement of any one
qubit (as one party) with the remaining qubits (as the other party).
Under such a bipartition, the pure state (\ref{arbitrary pure state})
above can always be decomposed into Schmidt form \cite{Schmidt,
Fedorov-Miklin-14, Ekert-Knight-95}, a sum of only two terms because
the singled-out qubit itself has only two states, i.e.,
\begin{equation} \label{Schmidt decompose}
|\Psi _{1,2,...,N}\rangle =\sum_{n=1}^{2}\sqrt{\lambda _{n}^{(j)}}
|f_{n}^{(j)}\rangle \otimes |g_{n}^{(j)}\rangle ,
\end{equation}
where $|f_{n}^{(j)}\rangle$\ and $|g_{n}^{(j)}\rangle$ are the
``information eigenstates" of the reduced density matrices of the
$j$-th qubit and the remaining $N-1$ qubits, from whose joint state
the Schmidt Theorem \cite{Schmidt} determines the unique
two-dimensional partner of the $j$-th qubit. The Schmidt coefficients
$\lambda_{1}^{(j)}$ and $\lambda_{2}^{(j)}$ are the corresponding
eigenvalues of the reduced density matrix for one party, i.e., the
single-qubit, and are the same for both parties.

The degree of entanglement between the $j$-th qubit and the remaining
$N-1$ qubits can be characterized in a number of ways, frequently by
the Schmidt weight $K_j$, which is given in \cite{K}: $1/K_{j} =
(\lambda_{1}^{(j)})^{2} + (\lambda_{2}^{(j)})^{2}$. For our purposes
an alternate normalized form is more useful:
\begin{equation} \label{YDef}
Y_{j} = 1-\sqrt{2/K_{j}-1},
\end{equation}
which can also be recognized as two times the smaller of
$\lam^{(j)}_1$ and $\lam^{(j)}_2$. Thus this entanglement monotone
satisfies $0 \le Y_{j} \le 1$, where 0 indicates complete
separability (zero entanglement), and 1 denotes maximal entanglement.

Our first main result for this entanglement measure is a compact
generic and symmetric entanglement inequality applying to all the
$Y_j$ values for the $N$ qubits:
\begin{equation} \label{N-additivity}
Y_{j} \leq  \sum_{k\neq j}Y_{k}.
\end{equation}
The proof of inequality (\ref{N-additivity}) is lengthy and is given
in the supplementary material \cite{Suppl.Matl.}. One notes that the
inequality takes a form that is available to other N-party
entanglement monotones, such as the von Neumann entropy and concurrence which are concave functions of $Y$. However,
most importantly, inequality (\ref{N-additivity}) is uniquely tight. In fact, the concavity of other entanglement monotones with respect to $Y$ already indicates a looser form of inequality than (\ref{N-additivity}). By saying uniquely tight, it means that (\ref{N-additivity}) not only applies to all $N$-party qubit pure states, but additionally that those states exhaust the inequality, occupying
its interior and also its boundaries. We will develop this point in the
following sections. An interesting trivial consequence is obtained
immediately by adding $Y_j$ to both sides of (\ref{N-additivity}), to
get the total $Y_T$ of all one-party marginal entanglements on the
right side. Thus no arbitrarily chosen $j$-th member of the party
can capture more than half of the total $Y_T$. This is independent of
the way the $N$-body state is specified.  \\

\begin{figure}[!t]
\includegraphics[width=8cm]{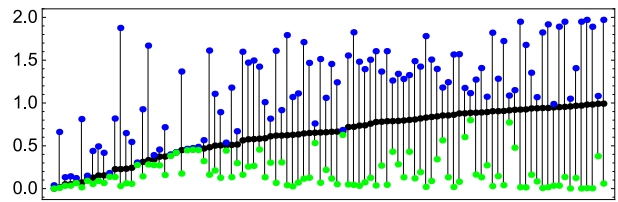}
\caption{Illustration of upper and lower bounds of $Y_1$. Black dots
indicate the $Y_1$ values for 100 randomly chosen 3-qubit pure
states. The blue (upper) dots identify each state's upper bound
determined from (\ref{N-additivity}) and the connected green (lower)
dots identify the lower bound obtained from (\ref{YtoC}) as converted
from the monogamy relation (\ref{concur}).}
\label{bound}
\end{figure}

\noindent {\bf Connection to Monogamy:} The constraints provided by
our inequality (\ref{N-additivity}) are different from those of the
well-known monogamy relations \cite{Coffman-etal-00,
Osborne-Verstraete-06}. However, a direct comparison can be made with
the $j$-th concurrence, the one analogous to $Y_j$, i.e., based on
the same bipartitioning. The Osborne-Verstraete proof
\cite{Osborne-Verstraete-06} of the $N$-qubit version of the
concurrence relation reads
\beq \label{concur}
C^2_{j \{N-1\}} \geq \sum_{k \neq j} C^2_{jk}.
\eeq
Here we denote $C_{jk}$ as the concurrence of parties $j$ and $k$
and, as in (\ref{N-additivity}), the right-hand sum includes the other $N-1$
qubits entangled with qubit $j$. On the left side of the inequality
$C_{j \{N-1\}}$ denotes the concurrence of $j$ with the other $N-1$
qubits which, taken together, have been reduced to single-qubit form
by the Schmidt Theorem \cite{Schmidt}. When applied to arbitrary
$N$-qubit pure states, $C_{j\{N-1\}}$ is a simple concave function of $Y_j$ in [0,1]:
\beq \label{C-Y}
C_{j\{N-1\}}^{2} = Y_{j}(2-Y_{j}).
\eeq
This converts to an inequality by substitution from (\ref{concur}) and we find
\begin{equation} \label{YtoC}
Y_{j} \geq  1-\sqrt{1- \sum_{k \neq j} C^2_{jk}}.
\end{equation}
This specifies a lower bound for $Y_j$ to accompany the upper bound
$\sum_{k\neq j} Y_{k}$ that is provided by our basic inequality
(\ref{N-additivity}). To give a concrete view of the new result,
these upper and lower bounds are shown in Fig.~\ref{bound} for 100
random three-qubit pure states. Notice that for some states the upper
and lower bounds almost coincide, indicating a similar tightness of
the two opposite bounds. A more precise upper bound of $Y_j$ would be
${\rm Min} [1,\sum_{k\neq j}Y_{k}]$ since all $Y_{k}$ are less than
1. This will simply bring all the higher upper bound points down to
1.\\

\begin{figure}[t!]
\includegraphics[width=8 cm]{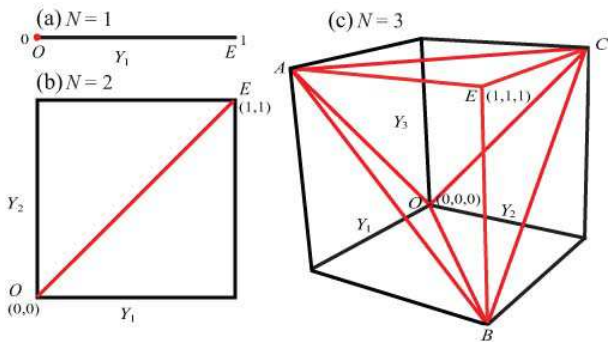}
\caption{$N$-dimensional spaces in which the vector ${\bf Y}$ is
defined, corresponding to (a) $N=1$ (unit line segment), (b) $N=2$
(unit square), and (c) $N=3$ (unit cube). In all cases, the point $O$
corresponds to no entanglement and the point $E$ to maximal
entanglement.}
\label{N123}
\end{figure}

\noindent{\bf Polytope Analysis:}  We now present our second main
result. We first exploit the $N$ different $Y_{j}$ measures by using them
to identify axes in a unit $N$-dimensional space. All possible
$N$-dimensional vectors ${\bf Y} = (Y_{1},Y_{2},...,Y_{N})$ live
inside a unit $N$-dimensional hypercube, since $0\le Y_{j}\le1$. For
$N=3$, for example, ${\bf Y}$ is a three-dimensional vector inside
the unit cube, shown in Fig.~\ref{N123}(c). The cube's origin
$(0,0,0)$ represents zero entanglement (corresponding to completely
separable states), while the opposite corner $(1,1,1)$ represents
maximal entanglement and corresponds to a GHZ state \cite{GHZ-07}. It
is worth stressing that ${\bf Y}$ is invariant to unitary local
transformations of the state. The new inequality (\ref{N-additivity}) implies that the region
inhabitable by the vectors ${\bf Y}$ (for pure states) is more
restricted than the $N$-dimensional unit hypercube.

Our second result is that these relations define a polytope, a hypervolume that is compact inside the unit
hypercube, where all possible ${\bf Y}$ are located. This polytope
geometrically represents the entanglement inequalities (\ref{N-additivity}). Each
inequality excludes a rectangular simplex whose hypervolume is given
by:
\begin{equation}
\prod_{j=1}^{N}\int_{0}^{1}[Y_{j}]^{j-1}dY_{j}=\frac{1}{N!}.
\end{equation}
Therefore the total available hypervolume, given the restrictions by
all $N$ such inequalities, is
\begin{equation}\label{availablehypervolume}
V_N=1-\frac1{(N-1)!}.
\end{equation}
To begin to visualize the effect of these constraints, we consider
the case $N=2$.

For $N=2$ there are two axes, $Y_{1}$ and $Y_{2}$, and their joint
range is the unit square shown in Fig.~\ref{N123}(b). In this case the
inequalities (\ref{N-additivity}) are simply $Y_{1} \geq Y_{2}$  and
$Y_{2} \geq Y_{1}$, or $Y_{1} = Y_{2}$. This restricts the allowed
region to a single line inside the unit square (dotted line in
Fig.~\ref{N123}(b)) running along the diagonal. On this line the
total entanglement $Y_T = Y_1 + Y_2$ runs from 0 to 2, from the
completely separable point $O$ to the maximally entangled point $E$
(corresponding to a Bell state). We note that for $N=2$ entanglement
can't be additively shared. The only way for $Y_1$ and $Y_2$ to add
up to any given $Y_T$ is $Y_1 = Y_2 = \frac{1}{2}Y_T$. Again, this agrees with
Eq.~(\ref{availablehypervolume}) -- the restricted volume fraction is
$V_2 = 0$, meaning that instead of an area, only a line is
inhabitable. The even simpler case $N=1$ restricts occupation to a
single point, $Y_1 = 0$.

\begin{figure}[!b]
\includegraphics[width = 8cm]{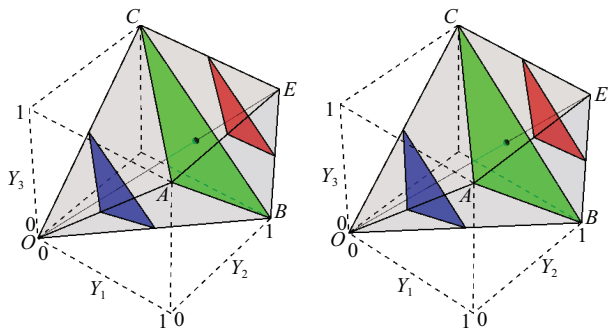}
\caption{Stereogram best viewed from a distance of about 25 cm. It
shows the inhabitable region of the 3D cube permitted by the
inequalities (\ref{N-additivity}), i.e., the polyhedron $OABCE$
(shaded in gray). See also Movie 1 \cite{Suppl.Matl.}. Also shown are
three triangular planar sections of this region transverse to the
unit cube's body diagonal. See also Movie 2 \cite{Suppl.Matl.}. }
\label{OnlyStereo}
\end{figure}

In the case $N=3$ additive sharing first comes into play. The three
entanglements $Y_{1},\ Y_{2},\ Y_{3}$ now reside inside a unit cube
(see Fig.~\ref{N123}(c)). From (\ref{N-additivity}) the generic three
qubit inequalities are given as
\begin{equation}
Y_{1}+Y_{2}\geq Y_{3}, \quad
Y_{2}+Y_{3}\geq Y_{1}, \quad
Y_{3}+Y_{1}\geq Y_{2}.\label{3inequality}
\end{equation}
One notes that when the three relations are equalities, each of them
defines a surface of the regular tetrahedron $OABC$. That is, the
three equilateral triangles $\triangle OAB$, $\triangle OBC$, and
$\triangle OCA$ are the surfaces separating allowed and forbidden
regions, as shown in Fig.~\ref{OnlyStereo}. Combining this with the
fact that $Y_1,Y_2,Y_3\le1$, the inhabitable region resulting from
the constraints by the three inequalities is simply the base-to-base
union of the regular tetrahedron $OABC$ and the rectangular
tetrahedron $ABCE$. This combined region is shown, shaded in gray, in
Fig.~\ref{OnlyStereo}. There are now many ways for the three
entaglements to add to a given total $Y_T$, and sharing is discussed
in the following section. As specified by
Eq.~(\ref{availablehypervolume}), the restricted volume is $V_3 =
1/2$. That is, only half of the cube is inhabitable by pure
three-qubit states.

Let us now view these results in regard to the well-known generalized
GHZ \cite{GHZ-07} and inequivalent W \cite{DVC-00} classes of three-qubit
states:
\beq \label{GHZ}
|\Psi_{\rm GHZ}\rangle = \cos\theta |0,0,0\ra + \sin\theta |1,1,1\ra,
\eeq
and
\beq \label{W}
|\Psi_{\rm W}\rangle = \alpha |1,0,0\ra + \beta|0,1,0\ra + \gamma|0,0,1\ra.
\eeq
It is straightforward to find that the GHZ states and their arbitrary
local unitary transformations live along the cube's body diagonal
line $OE$ (see Fig.~\ref{OnlyStereo}), according to
$Y_1=Y_2=Y_3=1-|\cos2\theta|$. The W class of states and their local
unitary transformations, on the other hand, live on the four surfaces
of the regular tetrahedron, i.e., $\triangle ABC$ as well as the inequality boundaries $\triangle OAB$, $\triangle OBC$,
and $\triangle OCA$. The occupation of these boundaries indicates the unique tightness of our inequalities (\ref{N-additivity}). The W class states all live away from the diagonal line $OE$ except for the three trivial cases when only
one of $\alpha,\beta,\gamma$ is nonzero, and one non-trivial case for
the perfectly symmetric W state when
$|\alpha|=|\beta|=|\gamma|=1/\sqrt{3}$. This state and the surface
$\triangle ABC$, which is the common base of the two tetrahedra, have
a special character that will be discussed in the following section.

For $N\ge3$, the inequalities are simultaneously all equalities only
if all the $Y_{j}$ vanish.

For $N \ge4$ qubits, the available region for all $Y_{j}$ is an
$N$-dimensional hypercube, and the allowed region, restricted by
Eq.~(\ref{N-additivity}), is an $N$-dimensional convex polytope.
According to Eq.~(\ref{availablehypervolume}), the ratio of the
allowed region $V_N$ to the unit hypervolume increases as the number
of qubits is increased, approaching unity as $N\to\infty$. This can
be understood as a consequence of multi-party entanglement because the
entanglement information shared among a higher number of parties is
less restricted than for a lower number.\\


\noindent{\bf Entanglement Additivity:} The geometric representation
provided by the ${\bf Y}$-space in the three-qubit case helps
visualize how entanglement inequalities provide a natural measure
of sharing or additivity, the freedom to distribute individual
entanglements in a way that the sum adds to a given $Y_T$. We start
by noticing that the domains of
different total entanglements $Y_T$ define triangles transverse to
the body diagonal (color triangles in Fig.~\ref{OnlyStereo}).
Inspection shows that the $Y_{T}$ value for these triangles varies
from 0 to 3, running from zero to maximal total entanglement. It is obvious that many combinations of the $Y_j$ are available to
sum to the total $Y_T$ in each transverse triangle.

\begin{figure}[!t]
\includegraphics[width = 4cm]{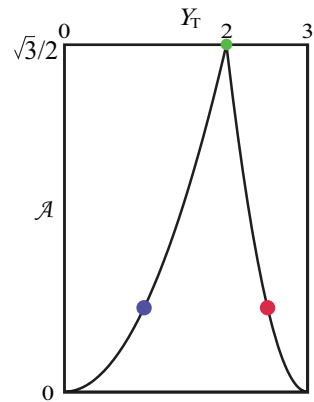}
\caption{ Additivity ${\cal A}$ is shown as a function of $Y_{\rm
T}$. The three colored dots correspond respectively to the three
colored triangles in Fig. \ref{OnlyStereo}.}
\label{OnlySharing}
\end{figure}

Our third main result is to adopt the area of each
triangle to serve as a natural measure of this entanglement additivity,
which we denote by ${\cal A}$. This interpretation is also natural
for $N=1$ and $N=2$. In those cases ${\cal A} = 0$ because the
counterparts of the transverse triangles are simply zero-area points,
corresponding to the lack of alternative arrangements of the
individual $Y_j$. The relation between ${\cal
A}$ and the amount of entanglement to be shared is not a linear
relation, but a piece-wise quadratic of the form:
\beqa
{\cal A} &=& \frac{\sqrt{3}}{2} \times \left\{\begin{array}{cc}
Y_{T}^{2}/4, & 0 \leq Y_{T} \leq 2,\\
(3-Y_{T})^{2}, & 2\leq Y_{T}\leq 3.
\end{array}\right.
\eeqa
The additivity ${\cal A}$ is graphed for $N = 3$ in Fig.~\ref{OnlySharing}, where we see that it is peaked
around its maximum of $\sqrt{3}/2$ at $Y_{\rm T}=2$, corresponding to
the triangle $\triangle ABC$. Clearly, more total entanglement $Y_T$ does not
guarantee greater additivity.

Note that the triangle joining the tetrahedral bases and shown in green
in Fig.~\ref{OnlyStereo}, has a special character. It
contains all of the  W states in Eq.~(\ref{W}) that satisfy ${\rm
Max} (|\alpha|^2,\ |\beta|^2,\  |\gamma|^2) \le1/2$; and the
perfectly symmetric W state lives at the point where the cube's body
diagonal $OE$ intersects the triangle. Thus W-like states can exhibit
maximum additive entanglement sharing. At the apexes
$O$ and $E$, corresponding to $Y_{\rm T}=0$ and $Y_{\rm T}=3$,
respectively, the transverse triangles have zero area. This is
intuitively correct since neither zero entanglement nor full total
entanglement can be distributed in any different way. In contrast to
the W state, one also notes that for any given $Y_T$, each GHZ-like state is a
single point on the body diagonal line $OE$ and so it permits
zero sharing (${\cal A}=0$).

One can easily extend the above analysis to the $N$-qubit case, where
additivity is then defined as the hyperarea of the
$(N-1)$-dimensional inhabitable polytope of fixed $Y_{\rm T}$, normal
to the line $OE$ within the $N$-dimensional polytope restricted by
the $N$ inequalities (\ref{N-additivity}). This expression is given
by a piecewise polynomial of $Y_{\rm T}$ of order $N-1$, which
vanishes at the endpoints $Y_{\rm T}=0$ (corresponding to point $O$)
and $Y_{\rm T}=N$ (corresponding to point $E$).\\


\noindent{\bf Summary:} Our results bring new light to the
understanding of quantum multiparty entanglement by focusing on the
simplest form of many-body bi-partitioning. This leads to the
emergence of the uniquely tight new quantum many-body inequality
(\ref{N-additivity}) that applies to each of the one-party marginal
entanglement monotones $Y_j$ defined in (\ref{YDef}), as proved in
\cite{Suppl.Matl.}. It provides a compact expression for the
restrictions that are acting, and they act in the opposite sense of
monogamy, as demonstrated in Fig. \ref{bound}. Thus they illuminate a
new aspect of generic resource sharing different from that
represented by monogamy.

Our inequalities provide an improved view of pure-state entanglement
and not only impose new quantum many-body limits on one-party
marginal entanglement, but also allow the sharing of entanglement
among members of an arbitrary many-body pure qubit state to be
quantified. They allow, we believe, the first quantitative definition
of additivity ${\cal A}$, a measure of the extent to which the
individual entanglements $Y_j$ can be added to produce a prescribed
total $Y_T$. One consequence is that more entanglement resource does
not necessarily mean greater additivity. This is obviously applicable
to tasks when optimum sharing instead of maximum entanglement is to be emphasized.

We have shown that all of the consequences of the inequality
(\ref{N-additivity}), and the character of the additivity measure
${\cal A}$, can be associated with the surfaces and volumes of
allowed vector spaces within hypercubes. These polytopes are
illustrated for $N=3$ in Fig. \ref{OnlyStereo}. The simplicity
of our approach is a key to our results. Preliminary numerical
results support the speculation that the same inequalities of $Y_j$
hold for pure states of many-body $M$-level systems, where the
normalized entanglement monotone becomes $Y_{j} = 1 -
\sqrt{\frac{M/K_{j}-1}{M-1}}$. This may supply a route for discovery
of additional resource-sharing equalities that may be based on
inequalities reported for quantum marginal multiparticle entanglement
by Walter, et al., \cite{Walter-etal-13} for higher dimensional
systems than qubits.

As will be discussed elsewhere, our entanglement inequality
(\ref{N-additivity}) for pure states remains relevant to mixed-state
generalizations. Then the inhabitable regions of the hypercube are
different. Dynamical trajectories within the allowed polytopes are
also under investigation, as well as alternative interpretations of
the ${\bf Y}$ space.\\


\noindent{\bf Acknowledgement:} We acknowledge partial financial
support from the National Science Foundation through awards
PHY-0855701, PHY-1068325, PHY-1203931, PHY-1505189, and INSPIRE PHY-1539859.


\end{document}